\documentclass[aps,prb,twocolumn,showpacs,preprintnumbers,amsmath,amssymb]{revtex4-1}

\usepackage{graphicx}
\usepackage{dcolumn}
\usepackage{bm}
\usepackage{color}

\usepackage{ulem}

\newcommand{\bra}[1]{\langle #1|}
\newcommand{\ket}[1]{|#1\rangle}

\newcommand{\be}{\begin{equation}}
\newcommand{\ee}{\end{equation}}
\newcommand{\bea}{\begin{eqnarray}}
\newcommand{\eea}{\end{eqnarray}}

\newcommand{\eq}[1]{Eq.~(\ref{#1})}
\newcommand{\fig}[1]{Fig.~\ref{#1}}

\newcommand{\e}{\varepsilon}

\newcommand{\s}{\sigma}
\newcommand{\G}{\Gamma}
\newcommand{\up}{\uparrow}
\newcommand{\down}{\downarrow}

\begin{document}

\title{Spin-resolved Andreev transport through double-quantum-dot Cooper pair splitters}

\author{Piotr Trocha}
\email{ptrocha@amu.edu.pl}
\affiliation{Faculty of Physics, Adam
Mickiewicz University, 61-614 Pozna\'n, Poland}

\author{Ireneusz Weymann}
\email{weymann@amu.edu.pl}
\affiliation{Faculty of Physics, Adam
Mickiewicz University, 61-614 Pozna\'n, Poland}

\date{\today}

\begin{abstract}
We investigate the Andreev transport through double quantum dot Cooper pair splitters
with ferromagnetic leads. The analysis is performed with the aid of the real-time
diagrammatic technique in the sequential tunneling regime.
We study the dependence of the Andreev current, the differential conductance
and the tunnel magnetoresistance on various parameters of the model
in both the linear and nonlinear response regimes.
In particular, we analyze the spin-resolved transport
in the crossed Andreev reflection regime, where a blockade
of the current occurs due to enhanced occupation of the triplet state. We show
that in the triplet blockade finite intradot correlations can lead to considerable
leakage current due to direct Andreev reflection processes.
Furthermore, we find additional regimes of current suppression
resulting from enhanced occupation of singlet states,
which decreases the rate of crossed Andreev reflection.
We also study how the splitting of Andreev bound states,
triggered by either dot level detuning, finite hopping between the dots or magnetic field,
affects the Andreev current. While in the first two cases
the number of Andreev bound states is doubled, whereas
transport properties are qualitatively similar,
in the case of finite magnetic field further level splitting occurs,
leading to a nontrivial behavior of spin-resolved transport characteristics,
and especially that of tunneling magnetoresistance.
Finally, we discuss the entanglement fidelity
between split Cooper pair electrons and
show that by tuning the device parameters fidelity can reach unity.
\end{abstract}



\maketitle


\section{Introduction}


Quantum dots coupled to superconducting and normal leads
provide very promising systems to study the interplay between the
superconducting correlations and the mesoscopic electronic transport.
\cite{FranceschiNatNano2010,rodero11}
Such hybrid nanostructures have recently attracted a lot of attention
due to the possibility to control and split the Cooper pairs.
\cite{deutscherAPL00,beckmannPRL04,russoPRL05,hofstetterSC,herrmann10SC,
hofstetterPRL11,Das12,SchindelePRL12,FulopPRB14,TanPRL15}
When the quantum dot is attached to normal and superconducting lead and for voltages
smaller than the superconducting energy gap $\Delta$,
the current flows through the system due to Andreev reflection. \cite{andreev}
More specifically, transport occurs then through sub-gap, Andreev bound states (ABS),
which were recently probed experimentally using bias spectroscopy.
\cite{leePRL12,leeNatNano14,schindelePRB14,kumarPRB14}
For three-terminal systems, e.g. with one superconducting and two normal leads,
the Cooper pair, when leaving the superconductor,
can enter either to the same normal lead or can be split
when the two electrons forming Cooper pair end in different leads.
The former process is known as direct Andreev reflection (DAR),
whereas the latter one is referred to as crossed Andreev reflection (CAR).
Usually both processes contribute to the Andreev current,
however, under certain conditions, by properly changing device parameters, one can
tune the contributions due to CAR and DAR processes or even suppress one of them.
This can be obtained in the case when the normal leads are ferromagnetic.
Then, in the antiparallel magnetic configuration of the device,
for leads with large degree of spin polarization,
only CAR processes contribute, since each lead supports electrons of opposite spin.
\cite{weymannPRB14,trochaPRB14}

Another interesting system, which allows for controllable manipulation
of Cooper pairs, can be made of double quantum dots (DQDs).
Recent experiments have shown that DQDs can work as
Cooper pair beam splitters, whose operation can be controlled by gate voltages.
\cite{hofstetterSC,herrmann10SC,hofstetterPRL11,Das12,SchindelePRL12,FulopPRB14,TanPRL15}
In contrast to single quantum dot hybrid systems, the DQD setup
allows to study pure CAR transport regime by considering
suitable system's parameters. Specifically, in real double quantum dots
the intradot Coulomb repulsion can be much larger than other energy scales,
\cite{hofstetterSC,kellerNatPhys2014}
which for a wide range of applied bias voltages prevents double occupancy of each dot.
As a consequence, the direct Andreev reflection processes,
which require simultaneous transfer of two electrons with opposite spins
by the same dot, become suppressed.
However, Andreev reflection processes can still occur through CAR processes,
in which the electrons forming the Cooper pair are transferred
simultaneously through the two dots.
Another advantage of DQDs is the possibility of independent level tuning
of each individual dot. Due to this ability, it has been shown experimentally
that Cooper pair splitting can be dominant on resonance,
whereas out of resonance elastic cotunneling processes dominate.
\cite{hofstetterPRL11,kleineEPL09}

Transport properties of DQD Cooper pair beam splitters have
already been addressed in several publications,
\cite{eldridgePRB10,hiltscherPRB11,BursetPRB11,ChevallierPRB11,rechPRB12,cottetPRB12,
CottetPRL12,BrauneckerPRL13,CottetPRB14,ScherueblPRB14,Chen15}
which, among others, addressed the problem of coherence and entanglement
of split Cooper pairs and their probing,
\cite{BrauneckerPRL13,CottetPRB14,ScherueblPRB14,Chen15}
as well as the noise correlations
\cite{ChevallierPRB11,rechPRB12}
and Cooper pair microwave spectroscopy.
\cite{cottetPRB12,CottetPRL12}
These investigations were performed for DQD Cooper
pair beam splitters with nonmagnetic leads.
However, because using ferromagnetic leads can be important to estimate
entanglement between split electrons, \cite{klobusPRB14}
providing comprehensive study of transport properties of
DQD Cooper pair splitters with ferromagnetic contacts seems desirable.
The analysis of Andreev transport through such systems
is thus the goal of the present paper.
Furthermore, DQDs coupled to a superconductor and to two ferromagnetic contacts,
can exhibit a considerable tunnel magnetoresistance (TMR) and generate large spin current.
Therefore, such nanostructures are also interesting for spin nanoelectronics
and understanding their magnetoresistive properties is of great importance.
Transport properties of quantum dots attached to ferromagnetic
leads have already been broadly investigated both experimentally
\cite{sahoo,pasupathy,mentel,hamaya1,hamaya2,hamaya3,yang,hamaya4,
HauptmannNatPhys08,GaassPRL11}
and theoretically. \cite{rudzinski,braun,cottet,weymann,barnasJPCM08,trochaPRB09}
However, spin-resolved transport properties of hybrid dots, consisting of quantum dots coupled to
ferromagnetic and superconducting leads, have been mainly studied
in the case of single quantum dots,
\cite{feng03,cao,pengZhang,konig09,konigPRB10,csonka,wysokinskiJPCM,wojcikPRB14,bocian}
while the case of double quantum dots is largely unexplored. \cite{trochaPRB14,siqueiraPRB10}

\begin{figure}[t]
\includegraphics[width=0.9\columnwidth]{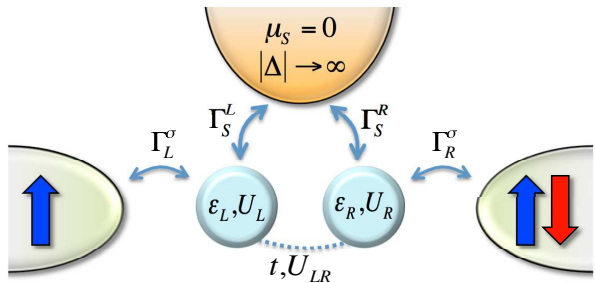}
  \caption{ \label{Fig:scheme}
  (Color online) Schematic of a double quantum dot Cooper pair splitter
  with ferromagnetic leads.
  The left (right) dot is coupled to the left (right) lead with the coupling strength
  $\G_{L}^\s$ ($\G_{R}^\s$) and each dot is coupled to a common superconducting lead
  with coupling $\G^S_L$ and $\G^S_R$ for the left and right dot, respectively.
  The magnetizations of the ferromagnetic leads can form
  either parallel or antiparallel magnetic configuration.
  The level energy and Coulomb repulsion in dot $i$
  are denoted by $\e_i$ and $U_i$,
  while $t$ and $U_{LR}$ describe the hopping
  and the Coulomb correlations between the two dots.}
\end{figure}

In this paper, we therefore investigate spin-dependent Andreev
transport through two single-level quantum dots, coupled
to one superconducting and two ferromagnetic leads.
Our analysis is performed with the aid of the
real-time diagrammatic technique in the lowest-order
expansion with respect to the coupling to ferromagnetic leads,
while the coupling to superconductor can be arbitrarily strong.
First, by assuming infinite correlations in the dots, we analyze
the pure CAR regime where the triplet blockade of the current occurs. \cite{eldridgePRB10}
We then show that even very large but finite intradot correlations
can lead to considerable leakage current in the triplet blockade.
We thoroughly study the behavior of the Andreev current, differential conductance
and TMR, deriving approximate zero-temperature formulas for TMR
in appropriate transport regimes. We also analyze
the effect of finite splitting of Andreev bound states on transport properties,
discussing the splitting caused by either level detuning, finite hopping between the dots
or finite magnetic field. For finite correlations in the dots,
we study transport properties in the full parameter space
and identify additional transport regimes where the current suppression occurs.
At the end, we also consider the entanglement fidelity
between split electrons forming Cooper pair and show
that, depending on parameters, fidelity can reach unity.

The paper is organized in the following way: Sec. II contains the description
of the DQD model and the method used in calculations.
The numerical results and their discussion in the crossed Andreev reflection regime
are presented in Sec. III. In Sec. IV
we analyze how the splitting of Andreev bound states affects
the transport properties. The next section is devoted
to the analysis of transport in the full parameter space
where both CAR and DAR processes are present.
The behavior of entanglement fidelity
on bias and gate voltages is studied in Sec. VI
and, finally, the paper is concluded in Sec. VII.


\section{Theoretical framework}


\subsection{Model Hamiltonian}

We consider double quantum dot Cooper pair splitter,
which is schematically displayed in Fig.~\ref{Fig:scheme}.
It consists of two single-level quantum dots,
each attached to its own ferromagnetic lead,
and both coupled to a common $s$-wave superconductor.
The magnetizations of ferromagnetic leads are assumed to form
either parallel (P) or antiparallel (AP) configuration.
Switching between these two configurations can be obtained upon
applying a small external magnetic field $B_s$.
We assume that this field is so small that it does not lead
to the splitting of the dot's level, neither affects it
the superconducting phase. The total system is modeled by the following effective Hamiltonian:
\begin{equation}\label{Eq:1}
  H=\sum_{\beta=L,R}H_{\beta}+H_S +H_{DQD}+H_T,
\end{equation}
where the first term, $H_{\beta}$, describes the left ($\beta=L$) and
right ($\beta=R$) ferromagnetic electrodes in the noninteracting quasiparticle approximation,
$H_{\beta}=\sum_{\textbf{k}\sigma}\varepsilon_{\textbf{k}\beta\sigma}
c_{\textbf{k}\beta\sigma}^{\dagger}c_{\textbf{k}\beta\sigma}$.
Here, $c_{\textbf{k}\beta\sigma}^{\dagger}$ ($c_{\textbf{k}\beta\sigma}$) is the creation (annihilation)
operator of an electron with the wave vector $\mathbf{k}$ and spin $\sigma$ in
the lead $\beta$, whereas $\varepsilon_{\textbf{k}\beta\sigma}$ denotes
the corresponding single-particle energy.
The second term in Eq.~(\ref{Eq:1}) describes
the $s$-wave BCS superconducting (S) lead
in the mean field approximation
\begin{equation}\label{Eq:2}
H_{S}=\sum_{{\mathbf
k}\sigma}  \varepsilon_{{\mathbf k}S\sigma}
     c^\dag_{{\mathbf k}S\sigma}c_{{\mathbf k}S\sigma}
     \!+\!
     \Delta \sum_{{\mathbf k}}\limits\left( c_{{\mathbf
k}S\downarrow}c_{{-\mathbf k}S\uparrow} + {\rm h.c.}\right)
\end{equation}
with $\varepsilon_{{\mathbf k}S\sigma}$ denoting the relevant
single-particle energy and $\Delta$ standing
for the order parameter of the superconductor,
which is assumed to be real and positive.
The third term of the Hamiltonian (\ref{Eq:1}) describes
the two single-level quantum dots and acquires the following form:
\begin{eqnarray}\label{Eq:3}
  H_{DQD}=\sum_{i=L,R}\left(\sum_{\sigma} \varepsilon_i d_{i\sigma}^{\dagger}d_{i\sigma} +B_zS_{iz}
  +U_i n_{i\uparrow}n_{i\downarrow}\right) \nonumber \\
   + \sum_{\sigma,\sigma'}\limits U_{LR} n_{L\sigma}n_{R\sigma'}
  + t \sum_\s (d_{L\s}^\dag d_{R\s} + d_{R\s}^\dag d_{L\s}), \;\;
\end{eqnarray}
where $d_{i\s}^\dag$ creates a spin-$\s$ electron in dot $i$
of energy $\varepsilon_{i}$, $n_{i\s} = d_{i\s}^\dag d_{i\s}$,
$U_L$ ($U_R$) is the Coulomb correlation energy of the left (right) dot,
and $B_z$ denotes external magnetic field in units of
$g\mu_B\equiv 1$ with $S_{iz} = (n_{i\uparrow} - n_{i\downarrow})/2$.
$U_{LR}$ and $t$ stand for the interdot Coulomb repulsion
and the hopping between the dots, respectively.

The last term of the Hamiltonian describes tunneling of electrons
between the leads ($L,R,S$) and the two dots
\begin{eqnarray}\label{Eq:4}
H_T=\sum_{\mathbf{k}\sigma}\limits\sum_{i=L,R}
   \limits (V_{\mathbf{k}\sigma}^i c^\dag_{\mathbf{k}i\sigma}d_{i\sigma}+\rm
   h.c.)
   \\ \nonumber
   +
   \sum_{\mathbf{k}\sigma}\limits\sum_{i=L,R}
   \limits (V_{i\mathbf{k}\sigma}^S c^\dag_{\mathbf{k}S\sigma}d_{i\sigma}+\rm
   h.c.) ,
\end{eqnarray}
with $V_{\mathbf{k}\sigma}^i$ ($V_{i\mathbf{k}\sigma}^S$), for $i=L,R$,
denoting the relevant tunneling matrix elements between the two dots
and ferromagnetic leads (the superconducting lead).
In the following, we assume that these matrix elements
are $\mathbf{k}$ and $\sigma$ independent,
$V_{\mathbf{k}\sigma}^i \equiv V^i$ and
$V_{i\mathbf{k}\sigma}^S \equiv V_i^S$.
The coupling of the dots to respective ferromagnetic leads
can be parametrized by, $\Gamma_{i}^{\sigma}=2\pi|V^{i}|^2\rho_{i}^\sigma$,
where $\rho_{i}^\sigma$ is the spin-dependent density of states of lead $i$.
Within the wide band approximation these couplings become
energy independent and constant.
Introducing the spin polarization of the $i$-th lead,
$p_i=(\rho_{i}^+-\rho_{i}^-)/(\rho_{i}^++\rho_{i}^-)$,
where $\rho_{i}^+$ ($\rho_{i}^-$) is the spin majority (minority) density of states,
the couplings can be written in the form,
$\Gamma_{i}^{\sigma} = (1+ \sigma p_i) \Gamma_i$,
with $\Gamma_i = (\Gamma_i^\up + \Gamma_i^\down)/2$.
Generally, each dot can be coupled to its lead
with different strength and the two leads can have different spin polarizations,
here, however, we restrict our analysis to symmetric systems
and note that the presented results are also qualitatively valid
for systems with weak asymmetry in the couplings.
We thus assume  $p_L = p_R \equiv p$ and $\Gamma_L = \Gamma_R \equiv \Gamma/2$.
Moreover, we also assume that the dots' levels are degenerate
$\e_L=\e_R\equiv \e$ and the dots' Coulomb energies are equal,
$U_L=U_R\equiv U$, unless stated otherwise.

Since in this paper we are only interested in Andreev transport, we can take the limit
of an infinite superconducting gap, $\Delta\rightarrow\infty$.
Then, the quantum dot system coupled to the superconducting lead
can be described by the effective Hamiltonian~\cite{rozhkov}
\begin{eqnarray}\label{Eq:5}
  H_{DQD}^{\rm eff}=H_{DQD} -\sum_{i=L,R}\frac{\Gamma_i^S}{2}\left(d_{i\uparrow}^{\dagger}d_{i\downarrow}^{\dagger}+ {\rm h.c.}\right)
  \\ \nonumber
  +\frac{\Gamma_{LR}^S}{2}\left(d_{R\uparrow}^{\dagger}d_{L\downarrow}^{\dagger} + d_{L\up}^{\dagger}d_{R\down}^{\dagger}+ {\rm h.c.}\right),
\end{eqnarray}
where $\Gamma_{LR}^S=\sqrt{\Gamma_{L}^S\Gamma_{R}^S}$.
The superconducting proximity effects are included in the last
two terms of Eq.~(\ref{Eq:5}). The first term describes local proximity effects
on each dot and arises due to direct Andreev reflection,
whereas the second term describes creation of nonlocal
entangled states between the two dots.
These nonlocal correlations are responsible for crossed Andreev reflection.
The effective pair potential $\Gamma_i^S$ ($i=L,R$)
is the coupling strength between the $i$-th dot and superconducting electrode
and acquires the form, $\Gamma_i^S=2\pi|V_{i}^{S}|^2\rho_{S}$,
where $\rho_{S}$ denotes the density of states of the superconductor in the normal state.
We assume that the couplings between the dots and superconductor
are equal, $\Gamma_L^S = \Gamma_R^S \equiv \Gamma_S$.

The device is biased in the following way:
The electrochemical potential of the superconducting lead is assumed to
be grounded, $\mu_S=0$, see \fig{Fig:scheme},
while the potentials of the left and right leads
are kept the same, $\mu_L=\mu_R \equiv \mu = eV$.
In this way the net current between the left
and right ferromagnetic lead is zero.
In the following, we use the convention that for positive bias,
$eV > 0$, the Cooper pairs tunnel to the superconductor,
while for negative bias, $eV < 0$, the Cooper pairs are extracted
from the superconducting electrode.

We would like to note that while the assumption of infinite
superconducting gap allows us to exclude normal tunneling processes
and study only the Andreev transport,
it needs to be taken with some care.
This is because in real systems the gap can be large, but is clearly finite.
\cite{nagamatsu01,heinrich13}
However, for relatively low bias voltages, as considered in this paper,
one can expect normal tunneling processes to be negligible and,
thus, the assumption of large superconducting energy gap is reasonable.

\subsection{Method}

In order to calculate the transport characteristics of the considered system, we employ
the real-time diagrammatic technique (RTDT),
\cite{schoeler,thielmann,palaNJP07,governalePRB08,weymannPRB08}
which is based on systematic perturbation expansion of the reduced density matrix and operators
of interest with respect to the coupling strength $\Gamma$.
Within the RTDT, in the stationary state the reduced density matrix $\hat{\rho}$
can be found from \cite{schoeler,thielmann}
\begin{equation}\label{Eq:master}
   \sum_{\chi'}W_{\chi,\chi'}P_{\chi'}=0 ,
\end{equation}
where the elements $W_{\chi,\chi'}$ of the self-energy matrix
$\mathbf{W}$ describe transitions between the states
$\ket{\chi}$ and $\ket{\chi'}$ on the Keldysh contour,
with $\ket{\chi}$ denoting the
many-body eigenstate of $H_{DQD}^{\rm eff}$, $H_{DQD}^{\rm eff}\ket{\chi} = \e_\chi\ket{\chi}$.
${\mathbf P}$ is the vector of diagonal density matrix elements
$P_{\chi}=\langle\chi|\hat{\rho}|\chi\rangle$,
which can be found from \eq{Eq:master} together with the normalization condition.
The current flowing from the ferromagnetic lead $i$
can be found from \cite{schoeler,thielmann}
\begin{equation}\label{Eq:8}
  I_{i}=\frac{e}{\hbar} {\rm Tr} \left\{ {\mathbf W}^{I_i}{\mathbf P} \right\},
\end{equation}
where ${\mathbf W}^{I_i}$ denotes the modified self-energy matrix ${\mathbf W}$,
which takes into account the number of electrons transferred through the junction $i$.

To find the occupation probabilities and the current,
we perform the perturbation expansion
of the self-energies, occupation probabilities and the current with respect
to the coupling strength to ferromagnetic leads $\Gamma$.
In our studies, we consider the weak coupling regime
and take into account only the first-order tunneling processes,
which correspond to sequential tunneling. We first determine the
self-energies using the respective diagrammatic rules~\cite{thielmann,konigPRB10}
and then calculate the occupation probabilities and the current by
using Eqs.~(\ref{Eq:master}) and (\ref{Eq:8}).

We also note that when $\Gamma\ll \Gamma_S$, as considered in the present paper,
and taking into account only the lowest-order tunneling processes
in the coupling to ferromagnetic leads, the reduced density matrix becomes
diagonal in the eigenbasis of the effective Hamiltonian.~\cite{eldridgePRB10}
This is why Eq.~(\ref{Eq:master}) includes only
the diagonal elements of the reduced density matrix.
Moreover, we would like to emphasize that while
the perturbation expansion with respect to the coupling strength $\Gamma$
is performed, no assumption on the strength of the Coulomb correlation parameters
is imposed, and they are treated in an exact way.
Furthermore, the assumption of the weak coupling regime implies that
the Kondo temperature of the system is exponentially small.
Thus, at temperatures considered in calculations,
the correlations leading to the Kondo effect are irrelevant and
do not need to be taken into account.
\cite{glazman89,avishaiPRB03,yeyatiPRM03}

\subsection{Quantities of interest}

The main quantity of interest is the current flowing through
the system due to Andreev reflection processes.
By calculating the currents $I_L$ and $I_R$ flowing through the left and right junctions,
the total current flowing into the superconductor can be
simply obtained from the Kirchhoff's law
\be \label{Eq:IS}
I_{S}=I_L+I_R,
\ee
together with the corresponding differential conductance, $G_{S} = dI_S/dV$.

Since the normal leads are ferromagnetic, the Andreev current
depends on the magnetic configuration of the device, which is assumed
to be either parallel or antiparallel.
We thus also calculate the tunnel magnetoresistance associated with the change of
magnetic configuration of the system, which is defined as
\cite{weymannPRB14}
\be \label{Eq:TMR}
   {\rm TMR}=\frac{I_S^{AP}-I_S^{P}}{I_S^{P}},
\ee
where $I_S^{P}$ and $I_S^{AP}$ denote the Andreev current
flowing into the superconductor in the parallel and
antiparallel magnetic configurations, respectively.
Note that this definition is opposite to that in the case of the Julliere model. \cite{julliere}
This is because for hybrid quantum dots with superconducting and
ferromagnetic leads, the Andreev current in the antiparallel configuration is usually
larger than that in the parallel configuration. \cite{weymannPRB14}

In the following we present and discuss the results on the Andreev transport through
DQD Cooper pair splitters with ferromagnetic leads obtained
within the sequential tunneling approximation.
We systematically study the behavior of the Andreev current,
the associated differential conductance and the TMR
in both the linear and nonlinear response regimes,
exploring basically the whole parameter space of the considered model.
In particular, to analyze the transport regime where CAR processes are dominant,
we first assume infinite Coulomb correlations on the dots,
so that DAR processes are totally suppressed.
We then relax this condition and allow for finite correlations in the dots
to study the transport properties basically in the whole parameter space.
We also analyze the effects of finite
hopping between the dots, nonzero detuning of DQD levels
and finite external magnetic field.
In addition, we calculate the entanglement fidelity
between the split electrons forming Cooper pair.

\section{Results in the crossed Andreev reflection regime}

When the charging energy of each dot is much larger than
Coulomb correlations between the dots and the applied bias voltage,
the rate of direct Andreev reflection is suppressed
and only CAR processes are possible.
This condition is in fact one of the main requirements
for the system to work as Cooper pair beam splitter.
\cite{hofstetterSC}
Therefore to analyze the Andreev transport in the case when DAR processes are absent
we now assume $U\rightarrow\infty$.
For the sake of simplicity of the following discussion,
let us also at this point neglect the hopping term assuming $t=0$
(the role of finite hopping will be considered further).

In the limit of infinite intradot correlations,
double occupation of each dot is forbidden
and only nine states (out of $16$) of the effective Hamiltonian (\ref{Eq:5})
are relevant for transport.
These are the following states: empty state $\ket{0,0}$,
four singly occupied states $\ket{\sigma,0}$, $\ket{0,\sigma}$,
and four doubly occupied states $\ket{\sigma,\sigma'}$ for $\sigma,\sigma'=\up,\down$,
where $\ket{\alpha,\beta}\equiv\ket{\alpha}_L\ket{\beta}_R$,
with $\ket{0}_i$ and $\ket{\s}_i$ denoting empty and singly occupied
states of dot $i$.
Due to the hopping and additional terms in \eq{Eq:5} resulting from the proximity effect,
the Hamiltonian is not diagonal in the above basis.
Diagonalizing the effective Hamiltonian, one finds a new basis
consisting of the following states: four singly occupied states
$\ket{\sigma,0}$, $\ket{0,\sigma}$, three triplet states $\ket{T_0}=(\ket{\!\down,\up}+\ket{\!\up,\down})/\sqrt{2}$,
$\ket{T_{\sigma}}=\ket{\sigma,\sigma}$, and two states
\begin{eqnarray}\label{Eq:7}
|\pm\rangle=\frac{1}{\sqrt{2}}\left(\sqrt{1\mp\frac{\delta}
{2\varepsilon_A}}|0,0\rangle\mp\sqrt{1\pm\frac{\delta}{2\varepsilon_A}}|S\rangle\right),
\end{eqnarray}
being linear combinations of empty state and singlet state,
$\ket{S}=(\ket{\!\!\down,\up}-\ket{\!\!\up,\down})/\sqrt{2}$.
We note that the triplet states become decoupled from
the superconductor since Cooper pairs
consist of two electrons with compensated spin, i.e., $S_{pair}=0$,
and there is no coupling between the singlet and triplet states.
The corresponding eigenenergies of the eigenstates given by \eq{Eq:7} are
$E_{\pm}=\delta/2\pm\varepsilon_A$,
where $\delta= 2 \varepsilon+U_{LR}$ denotes
detuning between the singlet states and the empty state,
while $2 \varepsilon_A = \sqrt{\delta^2+2\Gamma_{S}^2}$ measures
the energy difference between the states $|+\rangle$ and $|-\rangle$.

The Andreev bound state (ABS) energies are defined as~\cite{eldridgePRB10}
\begin{equation}\label{Eq:6}
E_{\alpha\beta}^{\rm ABS} = \alpha\frac{U_{LR}}{2} + \frac{\beta}{2}\sqrt{\delta^2+ 2 \Gamma_S^2} \;,
\end{equation}
where $\alpha,\beta=\pm$. These energies are the excitation energies
between doublet and singlet states of the double dot decoupled from the ferromagnetic leads.

\begin{figure}
\begin{center}
\includegraphics[width=0.8\columnwidth]{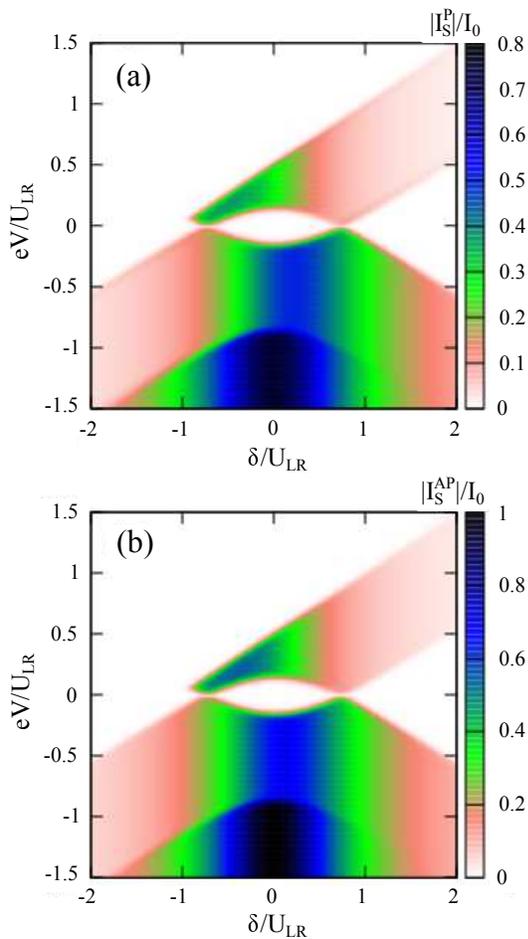}
  \caption{\label{Fig:2}
  (Color online) The absolute value of the Andreev current calculated
  for (a) the parallel ($I_{S}^{P}$) and (b) antiparallel ($I^{AP}_S$)
  magnetic configurations as a function of detuning $\delta = 2\e+U_{LR}$
  and the applied bias $eV$. The parameters are:
  $\G_S=0.5$, $T=0.015$, $\Gamma=0.01$, $t=0$, $B_z=0$, with $U_{LR}\equiv 1$ the energy unit,
  and $p=0.5$. The current is plotted in units $I_0=e\G/\hbar$.
  }
  \end{center}
\end{figure}

Due to the assumption, $U\rightarrow\infty$, double occupancy of each
dot is forbidden and direct Andreev tunneling becomes totally suppressed.
The only way to transfer charge between the double dot and the superconductor
is by crossed Andreev reflection, the process which involves two electrons
with opposite spins coming from different ferromagnetic leads.
For $eV > 0$, the two electrons tunnel to the superconductor,
while for $eV < 0$, the Cooper pairs are extracted
from the superconducting electrode and entangled pairs of electrons
are transmitted to ferromagnetic leads (each electron ends in different lead).

\subsection{Andreev current and differential conductance}

In Fig.~\ref{Fig:2} we show the dependence of the absolute value of the Andreev current
on the applied bias $eV$ and the detuning parameter $\delta=2\varepsilon +U_{LR}$
for the parallel and antiparallel magnetic configurations of the system.
At low bias, the Andreev current is generally suppressed due to the Coulomb blockade,
except for two values of $\delta$, $|\delta| = \sqrt{U_{LR}^2 -2 \Gamma_S^2}$,
where $E_{-+}^{\rm ABS} = E_{+-}^{\rm ABS} = 0$ and the corresponding
Andreev bound states are at resonance.
With increasing the bias, the current starts flowing once $|eV| > |E_{-+}^{\rm ABS}|$.
As the Andreev reflection becomes optimized for parameters corresponding to
particle-hole symmetry point, the Andreev current reaches maximum value
for small detuning, $\delta\approx 0$.
These two features are similar to those observed in a tree-terminal system
including a single quantum dot.~\cite{weymannPRB14}
However, in the present case the Andreev current exhibits
a striking difference, as it does not reveal the symmetry with respect to the bias reversal,
which has been present for the single dot system.
In the present case the absolute value of the Andreev current
reveals a very strong asymmetry with respect to the sign change of the bias.
As can be seen in Fig.~\ref{Fig:2}, for positive bias,
for which the double dot becomes occupied by two electrons,
the Andreev current ceases to flow.
In fact, for $eV>|(\delta+U_{LR})/2|$, the double dot is in the triplet state,
which explains the vanishing of the Andreev current,
as the symmetry of the triplet state does not match the symmetry of the s-wave superconductor.
The blockade region is therefore
independent of the magnetic configuration of the system
-- the Andreev current stops flowing due to the triplet blockade
in both the parallel and antiparallel alignments, see Fig.~\ref{Fig:2}.

\begin{figure}
\begin{center}
\includegraphics[width=0.8\columnwidth]{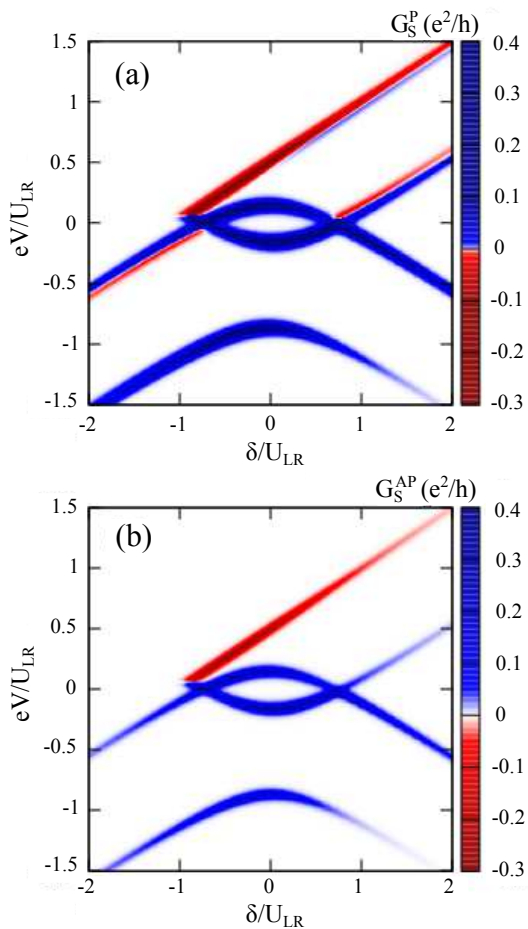}
  \caption{\label{Fig:3}
  (Color online) The differential conductance $G_S = dI_S/dV$
  of the Andreev current in the parallel ($G^{P}_S$) and antiparallel ($G^{AP}_S$)
  magnetic configurations as a function of detuning $\delta$ and applied bias voltage $eV$.
  Parameters are the same as in \fig{Fig:2}.}
  \end{center}
\end{figure}

Figure ~\ref{Fig:3} shows the dependence of the Andreev differential conductance
$G_S = dI_S / dV$ on the bias voltage $eV$ and detuning parameter $\delta$.
The sudden drop of the Andreev current around the bias voltage $eV\approx (\delta+U_{LR})/2$ for $\delta/U_{LR}>-1$
leads to the appearance of a pronounced negative differential conductance, see Fig.~\ref{Fig:3},
which is present in both magnetic configurations.
On the other hand, outside the triplet blockade the differential conductance
reveals positive peaks whenever the electrochemical potential
of ferromagnetic leads crosses one of the Andreev levels.

Except for the asymmetry due to the triplet blockade,
one can also notice another asymmetry of both
the current and differential conductance with respect to the bias reversal,
which is visible for large detuning $\delta$, see e.g. $\delta/U_{LR}=2$ in Figs.~\ref{Fig:2} and \ref{Fig:3}.
This asymmetry especially reveals in the differential conductance
when comparing the intensity of the low-bias peaks,
i.e. the peak associated with the Andreev level $E_{-+}^{\rm ABS}$ for
$eV>0$ (and $\delta>0$) with the amplitude of the maximum
associated with the level $E_{+-}^{\rm ABS}$ for $eV<0$ (and $\delta>0$),
see \fig{Fig:3}.
To understand this effect let us consider the differential conductance
for large value of detuning parameter $\delta/U_{LR}=2$.
The Andreev processes can occur in the system if the occupation
of states $\ket{+}$ and/or $\ket{-}$ is finite.
For large detuning the state $\ket{+}$ is
high in energy and does not play any role in the considered
bias voltage regime. The state which is relevant for Andreev transport
is the state $\ket{-}$.
One should also note that for large value of $\delta$ the state $\ket{-}$
contains only relatively small admixture of the singlet state, cf. Eq.~(\ref{Eq:7}),
while the empty state is mostly occupied.
This generally leads to small values of the current for large $\delta$.

For bias voltages such that $E_{+-}^{\rm ABS} <eV< E_{-+}^{\rm ABS}$, there
are no Andreev levels in the transport window and
the current flowing into/out of the superconductor is suppressed.
When $eV$ crosses one of those levels, either for positive or negative bias,
the current starts to flow and a peak appears in the differential conductance,
see Figs.~\ref{Fig:2} and \ref{Fig:3}.
It can be seen that the system becomes more transparent for $eV<0$,
when the Cooper pairs are extracted from the superconductor, than for $eV>0$,
when one injects electron pairs into superconducting lead.
This is because for positive bias voltage singly occupied
states become populated and, since
the singlet state is required for the current to flow into the superconductor,
the Andreev current is decreased.
More specifically, when passing the energy $\varepsilon$
(note that $\varepsilon=U_{LR}/2$ for $\delta = 2U_{LR}$)
the probability of finding the double dot in state $\ket{-}$
becomes strongly suppressed at the cost of enhanced occupation of one-electron states,
decreasing the Andreev current.
On the other hand, for negative bias voltage the singly occupied states play
little role in transport and the current is then larger compared to the case of $eV>0$.

\begin{figure}[t]
\includegraphics[width=0.8\columnwidth]{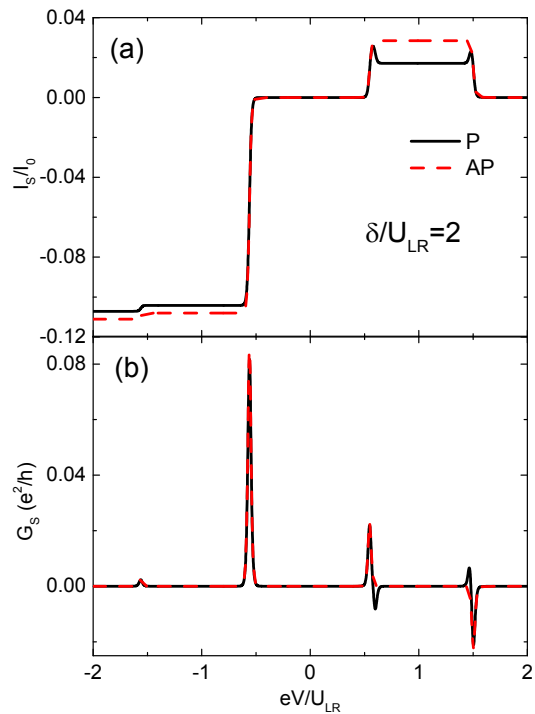}
  \caption{\label{Fig:6}
  (Color online) The Andreev current (a)
  in the parallel (solid line) and antiparallel (dashed line)
  magnetic configuration and the corresponding differential
  conductance (b) as a function of the bias voltage
  calculated for detuning parameter $\delta/U_{LR}=2$.
  The other parameters are the same as in \fig{Fig:2}.}
\end{figure}

For a better visualization of the behavior of the Andreev current
and related differential conductance in different transport regimes
we also plot the cross-sections of transport
characteristics for two values of detuning parameter $\delta$.
Figure~\ref{Fig:6} displays the bias dependence of the current
and differential conductance in both magnetic configurations
calculated for $\delta/U_{LR}=2$.
The current as a function of $eV$ exhibits well-defined steps corresponding
to consecutive Andreev bound states being active in transport,
while the differential conductance shows the respective peaks.
One can also see that the differential conductance
exhibits negative value due to the triplet blockade,
which occurs for bias voltage $eV\approx (\delta+U_{LR})/2$ and for $\delta/U_{LR}>-1$.
Interestingly, the differential conductance in the parallel
magnetic alignment reveals additional negative differential conductance,
which is not related to the triplet blockade,
see Figs.~\ref{Fig:3}(a) and \ref{Fig:6}(b).
This negative differential conductance develops
for $\delta>\sqrt{U_{LR}^2-2\Gamma_S^2}$
and for bias voltages $eV\approx E_{-+}^{\rm ABS}$.

The effect of negative differential conductance in the parallel configuration
can be explained bearing in mind that formation of
Cooper pairs involves two electrons with opposite spins.
In the parallel alignment there are more electrons
with one spin orientation than with the other one,
thus, the rate of electron pairs is determined by the
density of states of minority carriers. When the double dot
starts to be occupied by odd number of electrons
(here, singly occupied), the occupation probability of electrons with spin-up orientation
increases whereas that of electrons with spin-down decreases.
With further increase of the bias voltage
the occupation of the spin-up level becomes greatly enhanced,
whereas that of spin-down level becomes strongly suppressed,
giving rise to nonequilibrium spin accumulation.
As a consequence of spin accumulation,
the Andreev current becomes also suppressed,
which reveals as negative values in the differential conductance.
For reversed bias voltages the spin accumulation
becomes irrelevant since the double dot is in the state
$\ket{-}$, being rather empty with small admixture of singlet state.
Thus, Cooper pairs can be more easily extracted from the superconductor
compared to the opposite bias polarization.

\begin{figure}[t]
\includegraphics[width=0.8\columnwidth]{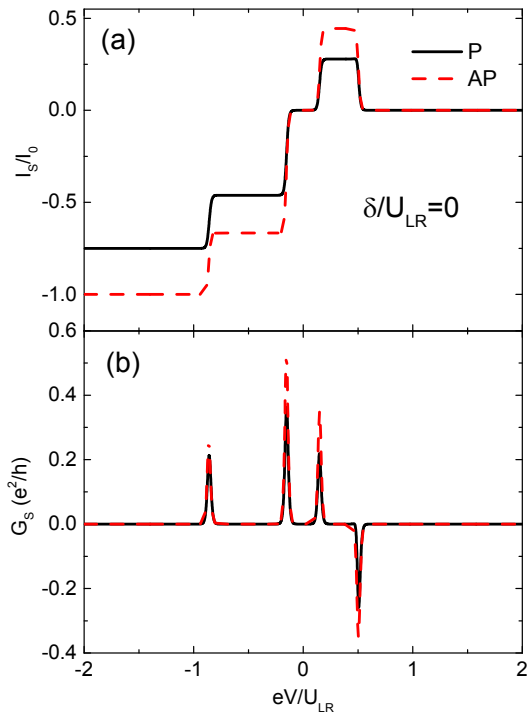}
  \caption{\label{Fig:7}
  (Color online) The same as in \fig{Fig:6} calculated for
  detuning parameter $\delta=0$.}
\end{figure}

The Andreev current and differential conductance as a function
of bias voltage in the absence of detuning are shown
in \fig{Fig:7}. Now, one can clearly see that while
for negative bias the current displays typical steps accompanied
with peaks in $dI_S/dV$, for positive bias voltage the current first increases
but then drops and becomes fully suppressed due to the triplet blockade,
see \fig{Fig:7}(a).
The associated negative differential conductance
is clearly visible in both magnetic configurations,
see \fig{Fig:7}(b).

\subsection{Tunnel magnetoresistance}

\begin{figure}[t]
\includegraphics[width=0.8\columnwidth]{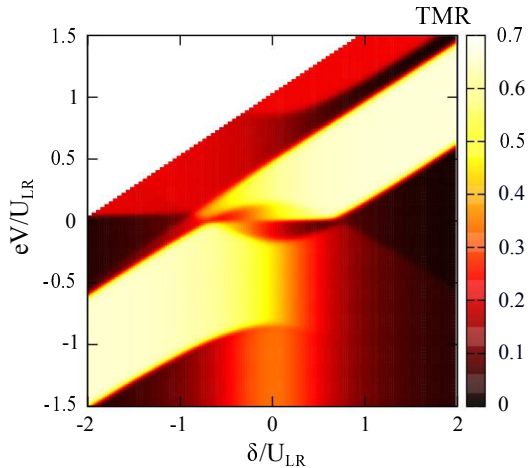}
  \caption{\label{Fig:5}
 (Color online) The tunnel magnetoresistance ${\rm TMR}$
  as a function of detuning $\delta$ and the bias voltage $eV$.
  The white region indicates the range of parameters
  where the TMR is undetermined since
  the current vanishes in both configurations due to the triplet blockade.
  The parameters are the same as in \fig{Fig:2}.}
\end{figure}

To observe more subtle differences between
the parallel and antiparallel magnetic configurations
one needs to use quantity which is more sensitive
to a change of magnetic alignment of ferromagnetic leads.
In Fig.~\ref{Fig:5} we present the dependence of TMR on
detuning $\delta$ and the bias voltage $eV$.
In the region determined by the equation $eV\gtrsim |(\delta+U_{LR})/2|$,
where the current ceases to flow due to the triplet blockade,
the TMR becomes indeterminate.
This region is marked by white area in Fig.~\ref{Fig:5}.
Moreover, the TMR is strongly suppressed for
$\delta>\sqrt{U_{LR}^2-2\Gamma_S^2}$ and
for the bias voltage $E_{+-}^{\rm ABS} <eV< E_{-+}^{\rm ABS}$.
For this transport regime however the first-order processes
are suppressed and to obtain correct value of TMR
higher-order tunneling events should be considered.
As has been shown recently,~\cite{weymannPRB14}
the cotunneling processes can lead to enhancement
of TMR in this transport regime.

Since the TMR takes well-defined values for parameters
corresponding to plateaus in the current, it is possible to find
some approximate analytical formulas for the TMR.
This can be done assuming very low temperatures
when the Fermi functions can be replaced by step functions.
The formula for TMR in the transport regime corresponding to $\delta/U_{LR}=2$
and describing the TMR at the plateau
$U_{LR}/2 \lesssim eV\lesssim 3U_{LR}/2$ is given by
\be \label{Eq:TMR1}
  {\rm TMR} = \frac{8(1+\varepsilon_A)}{(7+9\varepsilon_A)} \frac{2p^2}{1-p^2},
\ee
while the TMR for bias voltages $eV\lesssim -U_{LR}/2$ can be approximated by
\be
{\rm TMR} = \frac{2(\varepsilon_A-1)}{(3\varepsilon_A-1)} \frac{2p^2}{1-p^2}.
\ee
For $\delta/U_{LR}=2$, the first coefficient, $8(1+\varepsilon_A) / (7+9\varepsilon_A)$,
is very close to unity, while the second coefficient, $2(\varepsilon_A-1) / (3\varepsilon_A-1)$,
is close to zero. These two distinct values can be clearly seen in \fig{Fig:5}.
On the other hand, for $\delta=0$ and for negative bias voltage,
$-U_{LR}/2 - \Gamma_S/\sqrt{2}<eV< -U_{LR}/2 + \Gamma_S/\sqrt{2}$,
the TMR has a plateau of width $\sqrt{2}\Gamma_S$ and is given by
\be
{\rm TMR} = \frac{2}{3}\frac{2p^2}{1-p^2},
\ee
while for $eV < -U_{LR}/2 - \Gamma_S/\sqrt{2}$, the TMR reads
\be
{\rm TMR} = \frac{1}{2}\frac{2p^2}{1-p^2}.
\ee
For positive bias voltage and for $\delta=0$, the TMR
exhibits a plateau for
$U_{LR}/2 - \Gamma_S/\sqrt{2} < eV <  U_{LR}/2$
of width $\Gamma_S/\sqrt{2}$, at which it is given by
\be
{\rm TMR} = \frac{8}{9} \frac{2p^2}{1-p^2}.
\ee
Note that the TMR is always positive and smaller than
$2p^2/(1-p^2)$, \cite{weymannPRB14} see \fig{Fig:5}.

\subsection{The influence of intradot correlations}

\begin{figure}[t]
 \includegraphics[width=0.8\columnwidth]{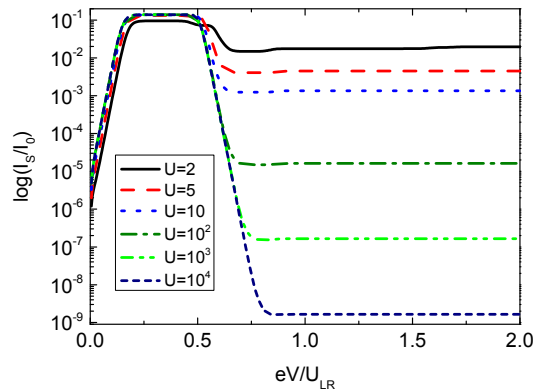}
  \caption{\label{Fig:logI}
  (Color online) The logarithm of the Andreev current
  in the parallel configuration for $\delta=0$
  as a function of the bias voltage for different
  Coulomb correlations $U$, as indicated.
  The other parameters are the same as in \fig{Fig:2}.}
\end{figure}

Results presented in previous sections were obtained in the limit of infinite
Coulomb correlations in the dots, so that
the Andreev current was mediated only by CAR processes.
We now relax this condition and allow for finite intradot Coulomb correlations,
and study their influence on the Andreev current and the TMR,
focusing on the triplet blockade regime.
Finite Coulomb correlations allow for nonzero current due to DAR processes,
which can lead to a nonzero leakage current in the triplet blockade.
Thus, once the current is finite, one can analyze the behavior of the TMR,
which is now well defined in the whole range of considered bias voltage.

Before proceeding with the discussion of the TMR,
in \fig{Fig:logI} we first study the bias dependence
of the Andreev current for zero detuning $\delta=0$.
To elucidate the role of finite intradot correlations
the current is plotted in the logarithmic scale.
This figure clearly shows how finite
Coulomb correlations affect the current in the triplet blockade regime.
Since the dependence of the current is qualitatively similar
in both magnetic configurations, we consider only the case of parallel
alignment. One can see that in the (unphysical) case
of infinite correlations, the current for $eV>U_{LR}/2$
is suppressed in an exponential way, $I_S\propto {\rm exp}(-eV/T)$.
However, even relatively large values of $U$
lead to finite current in the triplet blockade.
For semiconductor double quantum dots
the interdot correlations are typically an order of magnitude smaller
than the intradot correlations.~\cite{kellerNatPhys2014}
Although for recently-implemented
Cooper pair splitters based on nanowire DQDs
the capacitive coupling between the dots is even smaller,~\cite{hofstetterSC,hofstetterPRL11}
it can still play a role. As can be seen in \fig{Fig:logI},
the influence of direct Andreev reflection
on the triplet blockade is clearly nontrivial, for experimentally relevant parameters
it leads to relatively large leakage current in the triplet blockade,
see e.g. the curves for $U/U_{LR}\lesssim 10$.

\begin{figure}[t]
\includegraphics[width=0.8\columnwidth]{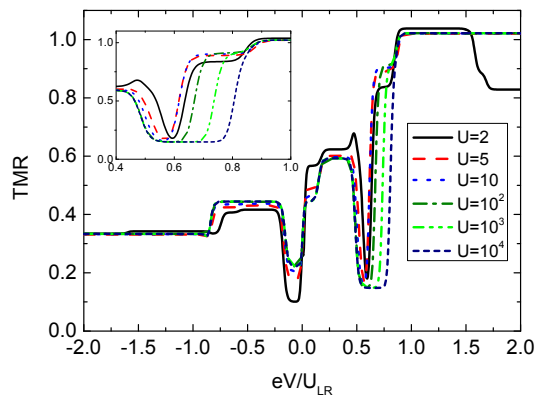}
  \caption{\label{Fig:TMRU}
  (Color online) The tunnel magnetoresistance
  as a function of the bias voltage calculated
  for $\delta=0$ and for different values of
  Coulomb correlations $U$, as indicated.
  The other parameters are the same as in \fig{Fig:2}.}
\end{figure}

The dependence of the TMR on the bias voltage
is presented in \fig{Fig:TMRU} for $\delta=0$
and for different values of intradot Coulomb correlations.
First of all, one can see that the behavior of TMR
for $eV<U_{LR}/2$ only very weakly depends on $U$.
Since the values of the TMR in this transport regime
were discussed in the previous section, let us only focus
on the range of bias voltages, $eV>U_{LR}/2$, where the DQD
is occupied by the triplet state.
The TMR becomes then greatly enhanced and reaches
values exceeding $2p^2/(1-p^2)$. Such large values of
TMR indicate that for finite intradot correlations not only
the rate of DAR processes becomes considarable, but also that
of CAR processes increases. This can be simply
understood by realizing that with lowering $U$ the occupation
of the triplet state decreases at the cost of other
states of the DQD, so that finite current due to
both types of Andreev reflection processes can flow.
Since DAR processes are not sensitive to a change of magnetic configuration of the device
(the two electrons tunnel always to the same, either left or right, lead),
the TMR provides an indirect information about CAR processes in the system,
the rate of which is clearly dependent on magnetic configuration.

\section{Splitting Andreev bound states}

We now study how the transport properties
of the DQD Cooper pair splitters change when one allows for
finite splitting of Andreev bound states. Such splitting
can be induced in various ways,
e.g. by detuning the DQD levels,
allowing for hopping between the dots
\cite{BursetPRB11}
or applying finite magnetic field.
We again focus on the same parameter space as in previous
sections, i.e. on the transport regime where
mainly CAR processes are present.
To gain a deeper insight of how the ABS become
affected, let us consider only $9$ states of the DQD (the limit of $U\to\infty$).
The analytical formula for Andreev bound states' energies can be then written as
\begin{equation}\label{Eq:17}
E_{\alpha\beta\gamma\delta}^{\rm ABS} = E_{\alpha\beta}^{\rm ABS} +
\frac{\gamma}{2}\sqrt{4t^2+(\Delta\e)^2}+\delta\frac{B_z}{2} \;,
\end{equation}
where $\Delta\e = \e_L - \e_R$ denotes the detuning of DQD levels,
$\gamma,\delta=\pm$,
and $E_{\alpha\beta}^{\rm ABS}$ is given by \eq{Eq:6}. Note that the splitting of ABS
results only from the splitting of single-electron states,
while the states $\ket{+}$ and $\ket{-}$ are not affected.
Moreover, one can also notice that finite level detuning
$\Delta\e$ can have a similar effect as finite hopping
between the dots when $\Delta\e = 2t$.
In the presence of either level detuning or hopping
each of the four ABS states splits into two,
which results in eight Andreev bound states. When external
magnetic field is additionally present,
these states split again and there are sixteen ABS states.
One can thus expect that the consequences of
ABS splitting will reveal as nontrivial features in transport characteristics.
In the following we study the spin-resolved transport properties
for finite detuning, hopping and, finally, for finite magnetic field.

\begin{figure}[t!]
\includegraphics[height=2\columnwidth]{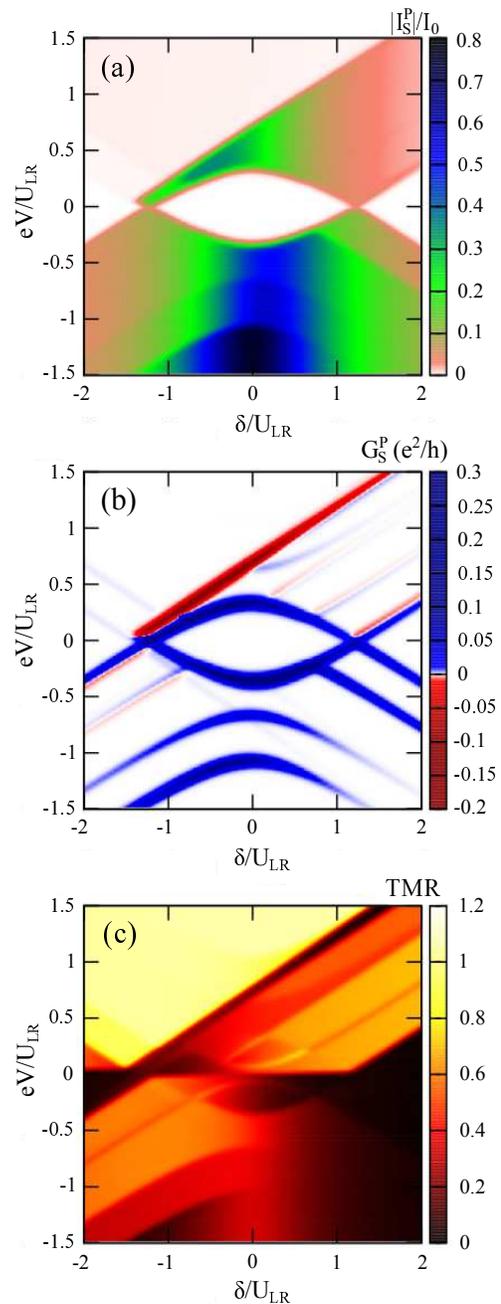}
  \caption{\label{Fig:DeltaE2D}
  (Color online) The bias voltage and detuning dependence
  of (a) absolute value of the Andreev current
  and (b) the respective differential conductance
  in the parallel magnetic configuration
  as well as (c) the TMR calculated for
  $\Delta\varepsilon/U_{LR} = 0.4$.
  The parameters are the same as in \fig{Fig:2} with $U / U_{LR}=10$.}
\end{figure}

\begin{figure}[t!]
\includegraphics[height=2\columnwidth]{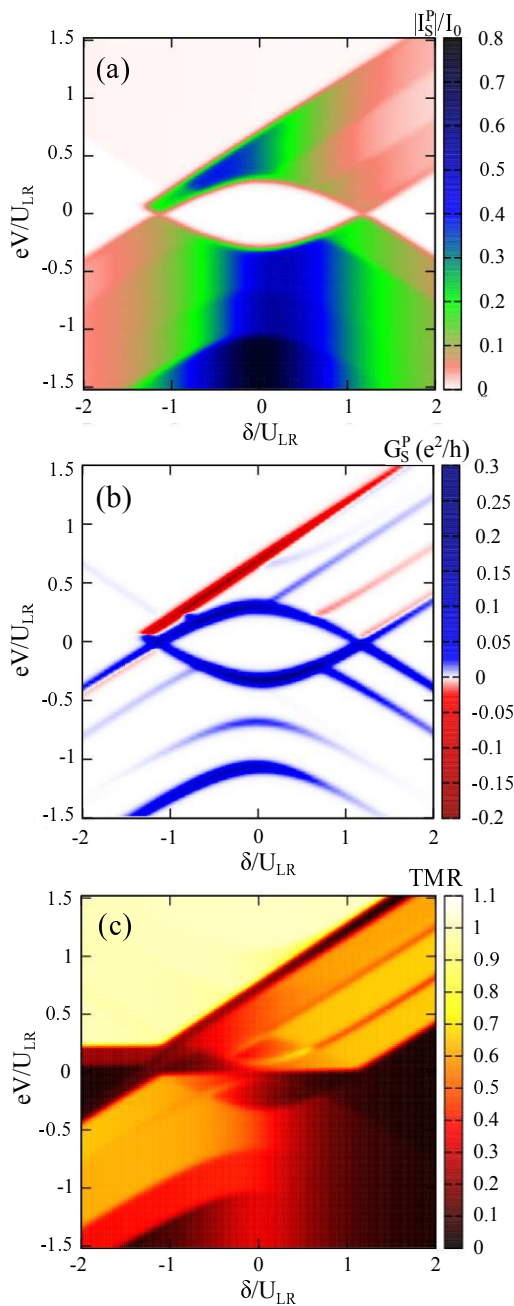}
  \caption{\label{Fig:hop2D}
  (Color online) The bias voltage and detuning dependence
  of (a) absolute value of the Andreev current
  and (b) the respective differential conductance
  in the parallel magnetic configuration
  as well as (c) the TMR
  calculated in the presence of finite hopping between the dots $t / U_{LR} = 0.2$.
  The other parameters are the same as in \fig{Fig:2} with $U / U_{LR}=10$.}
\end{figure}

Although we focus on transport regime where
doubly occupied states play negligible role,
in calculations we assume large but finite intradot
correlations to be able to determine the detuning and bias voltage
dependence of the TMR in the considered parameter space.
We thus assume $U / U_{LR} = 10$, if not stated otherwise.

\subsection{Finite level detuning or hopping}

From \eq{Eq:17}, one could simply expect finite $t$ and $\Delta\e$
to have the same effect on transport properties.
This is however not entirely true, as we show in the following.
The main difference results form the bonding and anti-bonding
states that form in the case of finite $t$ and are absent
if the splitting of ABS is caused only by detuning of the DQD levels.
The Andreev current, related differential conductance in the parallel configuration
and the TMR are shown in \fig{Fig:DeltaE2D} in the case of finite $\Delta\e$
and in \fig{Fig:hop2D} in the case of $t = \Delta\e/2$.
The fact that $t = \Delta\e/2$ guarantees
that the ABS excitation patterns occur
for similar bias voltages and detunings in both cases.

First of all, one can see that the number of
peaks in differential conductance has doubled
due to the two-fold splitting of ABS states.
Because of that, the current as a function of the bias voltage
for given detuning $\delta$ exhibits more Coulomb steps
compared to the case in the absence of ABS splitting.
The region of the triplet blockade can be clearly visible
for both finite $\Delta\e$ and $t$, see Figs.~\ref{Fig:DeltaE2D}(a) and \ref{Fig:hop2D}(a).
The main difference with the case shown in
\fig{Fig:3} (absence of splitting) is the shift of the triplet line
in the $(eV,\delta)$-plane by a factor of the induced
ABS splitting towards larger bias voltages.
Moreover, the negative differential conductance
for $\delta/U_{LR} \gtrsim 1$ associated with spin accumulation
in the doublet states is now also split, and one finds
additional regions of negative $G_S^{P}$,
see Figs.~\ref{Fig:DeltaE2D}(b) and \ref{Fig:hop2D}(c).
Note that for $\delta\gtrsim 1$, in the case of finite $\Delta/U_{LR}\e$,
there are four regions of negative differential conductance,
while for finite hopping $t$, there are only three.
Similar asymmetry can be observed for negative
detuning $\delta/U_{LR}\lesssim -1$, where
for $\Delta\e\neq 0$, one finds two regions
of current suppression, which occur for $eV<0$,
while for $t\neq 0$ there is only one, cf. Figs.~\ref{Fig:DeltaE2D} and \ref{Fig:hop2D}.
However, the effect of negative differential conductance,
which is associated with spin accumulation in doublet states,
is not that spectacular as in the case of the triplet blockade,
where the current suppression is much more pronounced
(note the nonlinear color scale used in Figs.~\ref{Fig:DeltaE2D} and \ref{Fig:hop2D}).

Although there are small differences
in the behavior of the current and differential conductance in the case
of nonzero $\Delta\e$ and $t$, the behavior of the TMR
is essential the same in both cases, see Figs.~\ref{Fig:DeltaE2D}(c) and \ref{Fig:hop2D}(c).
One can observe a large TMR in the triplet blockade, ${\rm TMR > 1}$,
while in the case when doublet states are relevant for transport
the TMR is close to ${\rm TMR} = 2p^2 / (1-p^2)$, cf. \fig{Fig:5}
and \eq{Eq:TMR1}. Now, however, this region is wider when changing the bias voltage
and the splitting of ABS reveals as stripes in TMR as a
function of the bias voltage and detuning,
which are most visible for $\delta \gtrsim 0$,
see Figs.~\ref{Fig:DeltaE2D}(c) and \ref{Fig:hop2D}(c).

\subsection{Finite magnetic field}

\begin{figure}[t]
\includegraphics[height=2\columnwidth]{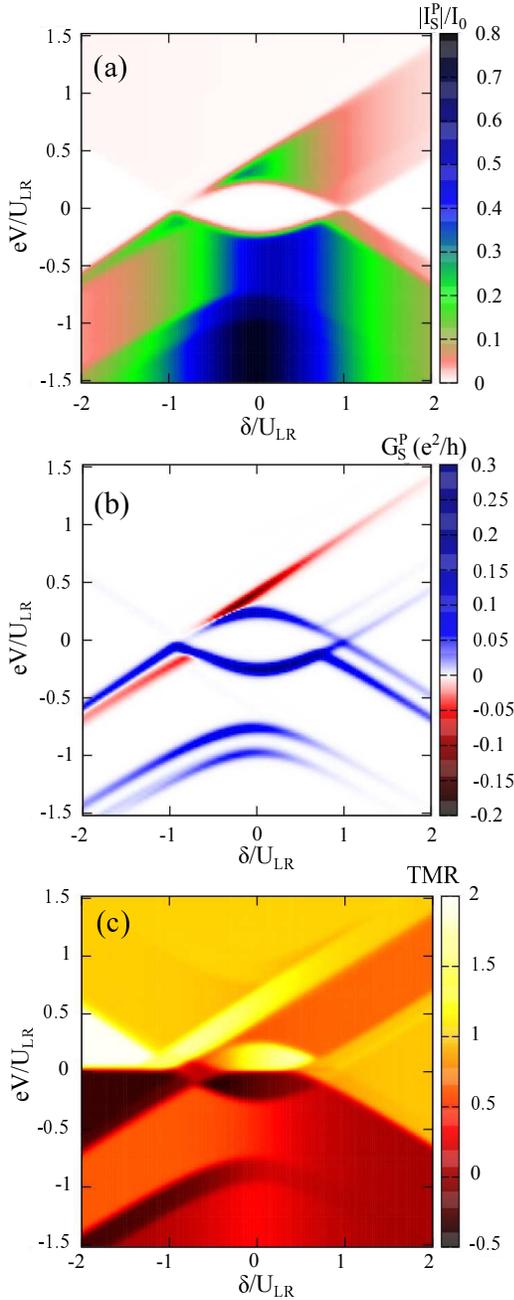}
  \caption{\label{Fig:B2D}
  (Color online) The bias voltage and detuning dependence
  of (a) absolute value of the Andreev current
  and (b) the respective differential conductance
  in the parallel magnetic configuration
  as well as (c) the TMR
  calculated in the presence of external magnetic field $B_z/U_{LR}=0.2$.
  The other parameters are the same as in \fig{Fig:2} with $U / U_{LR}=10$.}
\end{figure}

Here, we release the assumption about the smallness of external magnetic field $B_s$ needed to switch magnetic configuration of the ferromagnetic leads. Now, we assume that this field is strong enough to induce splitting of the ABS states, which can be achieved using ferromagnets with sufficiently large coercive field.
The splitting caused by external magnetic field has a larger
influence on spin-resolved transport compared to the splitting
due to either finite $\Delta\e$ or $t$.
The Andreev current, differential conductance in the parallel configuration
and the TMR for finite $B_z$ are shown in \fig{Fig:B2D}.
One can see that the triplet blockade region
is rather not affected by magnetic field, see \fig{Fig:B2D}(a).
This is due to the fact that while finite $B_z$ splits
the components of the triplet, the total occupancy
of all the triplet components does not change.
However, the finite magnetic field changes the range of bias voltage for which the triplet blockade occurs.
The negative differential conductance associated
with the triplet blockade is thus also clearly visible, see \fig{Fig:B2D}(b).
Interestingly, the negative differential conductance
due to the spin accumulation in the doublet states in now present only for
negative values of detuning $\delta/U_{LR}\lesssim-1$, cf. Figs. \ref{Fig:3} and \ref{Fig:B2D}.
This can be understood by realizing that
while for $\delta/U_{LR} \lesssim-1$ and $eV<0$, splitting caused by finite magnetic
field additionally enhances the spin accumulation,
in the case of $\delta/U_{LR} \gtrsim 1$ and $eV>0$,
magnetic field diminishes the spin accumulation.
Consequently, the current suppression
is reduced in the latter case,
while in the former one it is enhanced.

The bias voltage and detuning dependence of the TMR
is shown in \fig{Fig:B2D}(c). First of all, one can notice
a strong asymmetry with respect to the bias reversal,
which is most visible for small $eV$ and $\delta/U_{LR} \lesssim 1$.
For $\delta/U_{LR}\gtrsim 1$ and for low bias,
the DQD is empty and the TMR is positive in this transport regime,
and rather symmetric around $eV=0$.
On the other hand, when $|\delta/U_{LR}| \lesssim 1$, the DQD is singly occupied and
the TMR reveals then a strong asymmetry with respect to
the bias reversal. For positive bias voltage, there is a large positive TMR,
while for negative bias, the TMR becomes negative.
Such asymmetry is associated with the splitting
of the doublet ground state of the DQD.
It strongly affects CAR processes in the antiparallel configuration,
since the Cooper pair electrons tunnel then to either majority
or minority spin bands of the ferromagnets.
Because for one bias polarization
the electron occupying the DQD is the majority-band electron,
while for the opposite bias polarization, this electron belongs
to the spin minority channel, it effectively leads
to large asymmetry of the flowing current with respect
to the bias reversal, which is most visible in the TMR.
The effect of sign change of the TMR can be even more pronounced
in the case of $\delta/U_{LR} \lesssim -1$, where
the DQD is occupied by two electrons.
For $eV>0$, one then finds
${\rm TMR} \approx 2$, while for $eV<0$,
${\rm TMR} \approx -1/2$.
The mechanism leading to this asymmetry
is similar to that described above.
Note, however, that in the blockade
regions the current, and thus the TMR,
can be still modified by cotunneling processes. \cite{weymannPRB14}

Out of the Coulomb blockade region,
the changes in TMR are not that spectacular,
however, there are still considerable differences
compared to the case of $B_z = 0$.
First of all, negative values of the TMR
can be observed for $eV\lesssim -U_{LR}$ and $\delta/U_{LR} \lesssim -1$.
Moreover, although in the triplet blockade regime, one observes a large positive TMR,
similarly as in the absence of magnetic field, cf. Figs. \ref{Fig:5} and
\fig{Fig:B2D}(c), for finite $B_z$
there is an additional region of an enhanced TMR.
It occurs for voltages around $eV \approx (\delta+U_{LR})/2$
and is associated with the splitting of the triplet components.
Such splitting increases the difference in the currents
in both magnetic configurations, yielding ${\rm TMR} > 1$.

\section{Results in the full parameter space}

\begin{figure}[t]
\includegraphics[height=2\columnwidth]{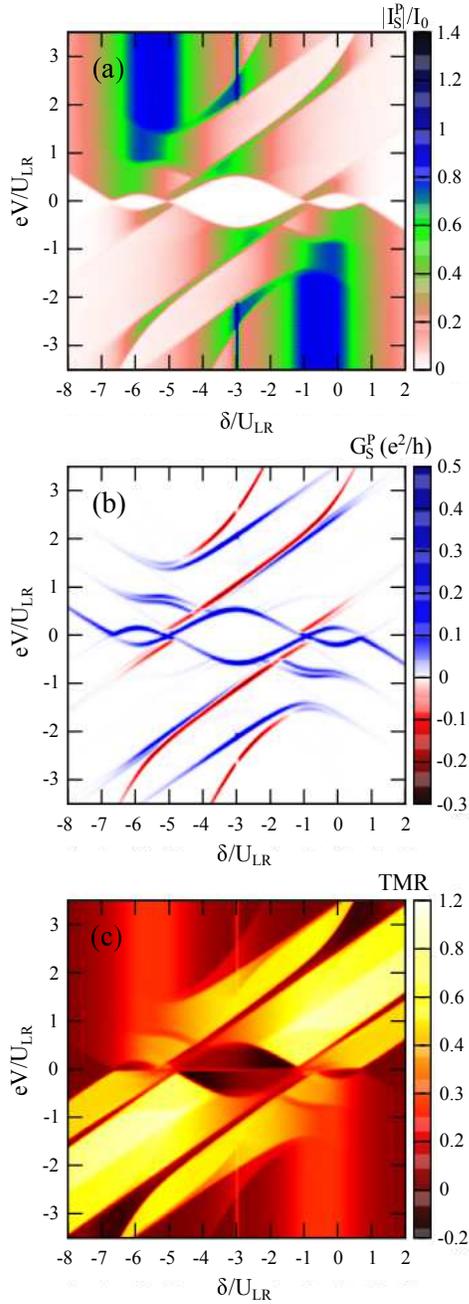}
  \caption{\label{Fig:full}
  (Color online) The absolute value of the Andreev current (a) calculated
  for the parallel ($I^{P}_S$)
  magnetic configuration, the corresponding differential
  conductance (b) and the tunnel magnetoresistance (c) as a function of detuning $\delta = 2\e+U_{LR}$
  and the applied bias $eV$. The parameters are
  the same as in \fig{Fig:2} with $U/U_{LR} = 2$.}
\end{figure}

To complete the analysis of Andreev transport through DQD Cooper pair
splitters, in this section we extend the discussion to the full parameter space.
Figure \ref{Fig:full} presents the bias voltage and detuning dependence
of the absolute value of the Andreev current and the corresponding differential conductance
in the parallel magnetic configuration as well as the TMR.
Transport characteristics are now symmetric with respect to the
particle-hole symmetry point of the DQD Hamiltonian,
$\e_{ph} = -U_{LR}-U/2$ ($\delta_{ph} = -U_{LR}-U$), with
immediate sign change of the bias voltage,
$I_S(eV,\delta>\delta_{ph}) = -I_S(-eV, \delta<\delta_{ph})$,
$G_S(eV,\delta>\delta_{ph}) = G_S(-eV, \delta<\delta_{ph})$,
and ${\rm TMR}(eV,\delta>\delta_{ph}) = {\rm TMR}(-eV, \delta<\delta_{ph})$, see \fig{Fig:full}.
Note that to enable direct comparison with previous results
we still plot transport characteristics as a function of $\delta = 2\e+U_{LR}$.
Moreover, we assumed relatively large capacitive coupling between the two
dots, $U=2U_{LR}$, to be able to show transport properties in the whole range of detuning $\delta$
in a single panel and not to obscure the features discussed previously,
which occur around $\delta \approx 0 $.
However, results are qualitatively the same for larger intradot correlations,
as checked numerically (not shown), the main difference
is in the distance between the resonances occurring now for $\delta/U_{LR}\approx-1$
and $\delta/U_{LR}\approx-5$, see \fig{Fig:full}, which increases with increasing  $U$.

Due to finite Coulomb correlations in the dots, there are
more Andreev states available for transport,
which generally reveals as steps in the bias dependence of the current
and corresponding peaks in the differential conductance.
One can clearly see the regime of the triplet blockade,
which occurs for $(\delta+U_{LR})/2 \lesssim eV \lesssim (\delta+U_{LR}+2U)$
and $\delta \gtrsim 0$ or $\delta \lesssim -2U-2U_{LR}.$
Note that there is a relatively large leakage current in the triplet blockade
due to finite intradot correlations, which allow DAR processes to participate in transport.
Interestingly, there are also another regions where the current
becomes suppressed and negative differential conductance occurs,
see Figs.~\ref{Fig:full}(a) and (b).
To present and discuss these effects in more detail,
let us show the relevant cross-sections of \fig{Fig:full}.
Since all transport features display an appropriate symmetry
with respect to $\delta = \delta_{ph}$, in the following
we will only analyze the results for $\delta\geq \delta_{ph}$.

\begin{figure}
\begin{center}
\includegraphics[width=0.8\columnwidth]{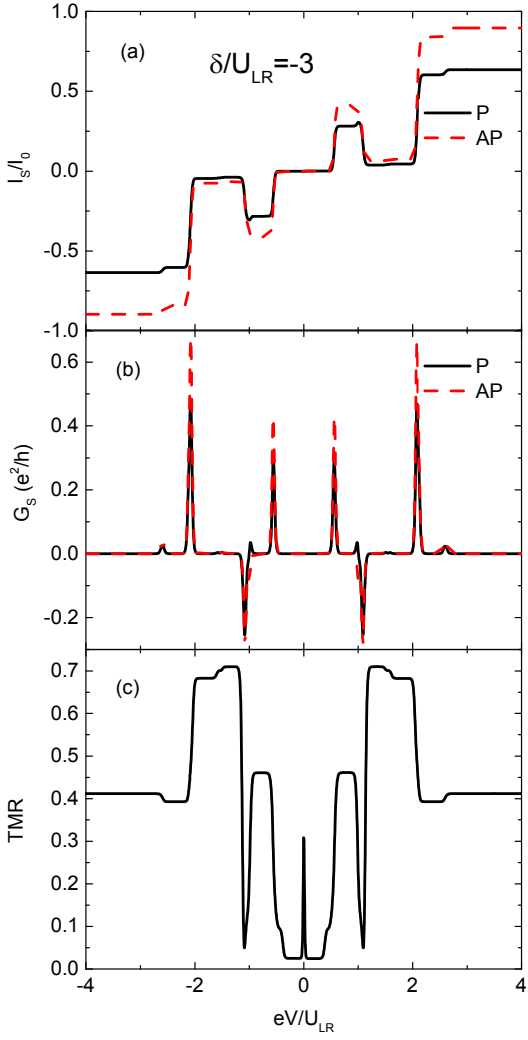}
  \caption{\label{Fig:full_dm3}
  (Color online) The Andreev current (a)
  and the corresponding differential conductance (b)
  in both magnetic configurations together
  with the tunnel magnetoresistance (c)
  as a function of applied bias $eV$ for $\delta/U_{LR}=-3$ $(\delta = \delta_{ph})$.
  The other parameters are the same as in \fig{Fig:full}.}
  \end{center}
\end{figure}

The current and differential conductance in both magnetic configurations
as well as the resulting TMR in the case of $\delta / U_{LR} = -3$
are shown in \fig{Fig:full_dm3}.
First of all, we note that transport characteristics
are now symmetric with respect to the bias reversal.
This is due to the fact that the presented data were obtained for the
particle-hole symmetry point of the model $\delta = \delta_{ph}$.
Moreover, one can see that the current does not increase
in a monotonic way, see \fig{Fig:full_dm3}(a). For $|eV| / U_{LR} \gtrsim 1$,
the Andreev current becomes suddenly suppressed
and the system exhibits a pronounced negative differential conductance,
see \fig{Fig:full_dm3}(b), which is present in both magnetic configurations.
The decrease of the current is related with an enhanced occupation of doublet states.
More precisely, for positive bias voltage $eV / U_{LR} \gtrsim 1$ and in the
antiparallel configuration, the occupation of the states
$(\ket{\!\up,d}-\ket{d,\up})/\sqrt{2}$ and $(\ket{\!\down,d}-\ket{d,\down})/\sqrt{2}$
is close to unity, which decreases the rate for both DAR and CAR Andreev processes.
Similar situation also holds for negative bias voltage,
but now the one-electron doublet states become occupied.
Consequently, the Andreev current becomes suppressed
and the system exhibit negative differential conductance.
However, with further increase of the bias voltage, $|eV| / U_{LR} \gtrsim 2$,
the occupation of the above doublet states decreases
and the current raises again, changing then monotonically with the bias voltage.

The above-described behavior is present in both magnetic configurations,
however, in the parallel configuration there is a strong
spin accumulation in the doublet states and the current is more
suppressed compared to the antiparallel configuration.
This is reflected in the behavior of the TMR on the bias voltage,
which is shown in \fig{Fig:full_dm3}(c).
At low voltage the TMR is negligible,
which is related to the fact that the system is occupied
by the singlet state $\alpha (\ket{0,0} - \ket{d,d}) + \beta (\ket{\!\up,\down} - \ket{\!\down,\up})$,
and thermally activated transport through this state
is insensitive to the change in magnetic configuration.
With increasing $eV$, the TMR increases at the first step in the current.
Interestingly, for voltages where the current is suppressed,
the TMR takes large values due to spin accumulation in the doublet states,
but then drops again with next step in the current
when the voltage is increased further on.

\begin{figure}
\begin{center}
\includegraphics[width=0.8\columnwidth]{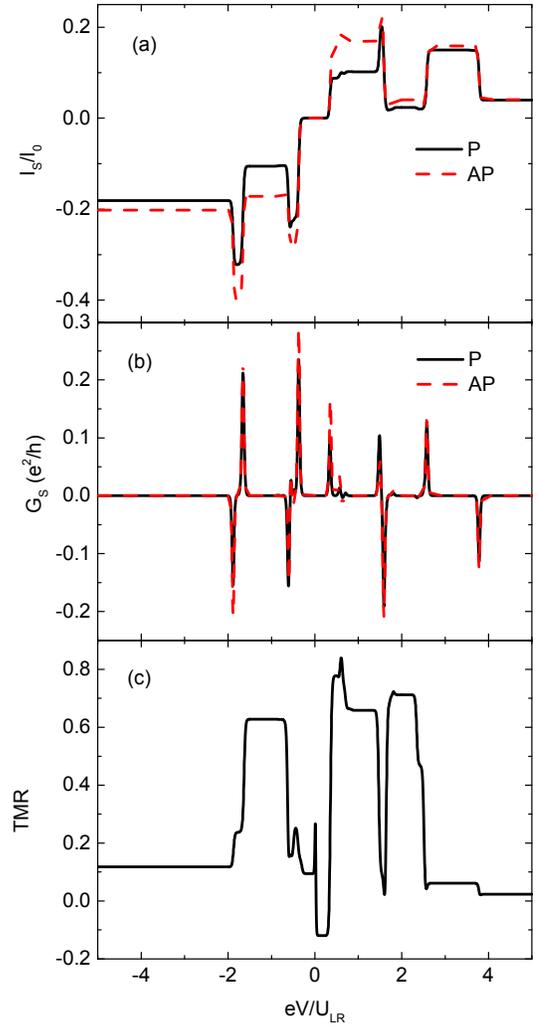}
  \caption{\label{Fig:full_dm2}
  (Color online) The same as in \fig{Fig:full_dm3}
  calculated for $\delta/U_{LR}=-2$.}
  \end{center}
\end{figure}

When moving away from the particle-hole symmetry point, $\delta = \delta_{ph}$,
there is a large change in the transport characteristics, see \fig{Fig:full}.
First of all, a pronounced asymmetry with respect to the sign
change of the bias voltage occurs. Moreover, transport characteristics
become more complex, since the number of negative differential conductance
regions increases and one can also find transport regimes where negative TMR occurs.
Here, let us discuss in somewhat greater detail the case of
$\delta/U_{LR} = -2$, which is presented in \fig{Fig:full_dm2}.

The Andreev current as a function of bias voltage is shown in \fig{Fig:full_dm2}(a).
Its bias dependence is not monotonic irrespective of bias polarization
and magnetic configuration of the device. For both positive and negative
bias voltage, there are two regions of current suppression
accompanied with respective negative differential conductance, see \fig{Fig:full_dm2}(b).
For positive bias, the current first decreases once $eV/U_{LR}\approx 3/2$,
which is associated with enhanced occupation of the doublet state
$[\ket{\!\up,d}-\ket{d,\up}]/\sqrt{2}$. In this transport regime
DAR processes are suppressed and transport is mainly due to CAR processes.
With increasing the bias voltage, $I_S$ increases for $eV/U_{LR}\approx 5/2$ to drop again once
$eV/U_{LR}\approx 4$, where the DQD becomes mainly occupied by the state $\ket{d,d}$.
Then, the rate of both Andreev reflection processes becomes decreased.
Similar features can be observed for negative bias voltage
and the mechanism leading to current suppression
and negative differential conductance is basically the same.
The first negative differential conductance occurs due to
enhanced occupation of the doublet state $[\ket{\!\down,0}-\ket{0,\down}]/\sqrt{2}$
(for $-1/2 \gtrsim eV/U_{LR}\gtrsim -3/2$),
while in the second suppression region the DQD is in the
state $\ket{0,0}$ (for $eV/U_{LR}\lesssim -2$).

Interestingly, in regions of current suppression due to enhanced
doublet occupation, the TMR exhibits rather large values, see \fig{Fig:full_dm2}(c).
This indirectly confirms that the main role is played by CAR processes,
which greatly depend on magnetic configuration of the system,
contrary to DAR processes. Consequently, one observes a large positive TMR effect.
On the other hand, in blockade regions due to enhanced occupation
of either empty or fully occupied DQD, the Andreev current
depends very weakly on the magnetic configuration of the system,
which implies that CAR processes play a minor role in transport.

One can also note that CAR processes become relevant
not only in blockade regions, but also for
$1/2\lesssim eV/U_{LR}\lesssim 3/2$, that is at the first plateau for positive bias voltage,
see \fig{Fig:full_dm2}(c). This is related with nonequilibrium spin accumulation
in the parallel configuration, due to which the occupation of the triplet component
$\ket{\!\up,\up}$ becomes enhanced and the Andreev current drops.
This triplet blockade is absent in the antiparallel configuration,
which leads to large difference in the currents in both configurations
and thus to large TMR effect.

Another interesting feature is the negative TMR, which occurs
at very low positive bias voltage, see \fig{Fig:full_dm2}(c).
In this transport regime the Andreev processes are suppressed
due to the fact that the DQD is occupied
by the doublet state $[\ket{\!\down,0}-\ket{0,\down}]/\sqrt{2}$
with almost unit probability. This occupation probability
is slightly lower in the parallel configuration for positive bias voltage
at the cost of small but finite occupation of the state
$\alpha[\ket{\!\up,0}+\ket{0,\up}]+\beta[\ket{\!\up,d}+\ket{d,\up}]$.
It is this state that allows for finite Andreev current
in the parallel configuration, giving rise to negative TMR effect.

\section{Entanglement fidelity}

In this section we study the entanglement fidelity between
split electrons forming a Cooper pair. Since the Cooper pair
is split and each electron tunnels to different arm
of the device in a CAR process, we again
focus on the transport regime where DAR processes
are excluded by assuming infinite intradot Coulomb correlations.

\begin{figure}
\includegraphics[width=0.8\columnwidth]{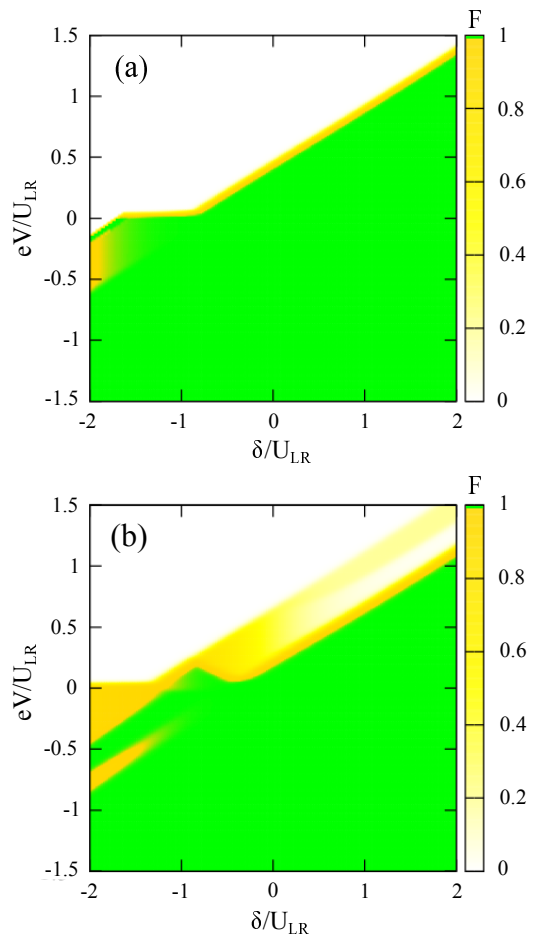}
  \caption{\label{Fig:F}
  (Color online) Fidelity $F$ in the parallel magnetic configuration
  as a function of detuning $\delta$ and the bias voltage $eV$
  calculated for (a) $\Delta\e = 0$
  and (b) $\Delta\e / U_{LR} = 0.4$.
  The dark (green) region corresponds to $F=1$.
  The other parameters are the same as in \fig{Fig:2}.}
\end{figure}

To analyze the fidelity let us consider
the Werner state,~\cite{werner} which has the following form,
\be\label{Eq:}
W(F)=F\ket{S}\bra{S}+(1-F)\frac{\mathbf{1} -\ket{S}\bra{S}}{3},
\ee
where $F$ denotes Werner fidelity.
For $F\leq 1/2$, the Werner state is unentangled,
whereas for $1/2<F\leq 1$, there exist purification protocols,
which can extract states with arbitrary large entanglement.
Werner fidelity for the considered system
is given by the formula,~\cite{legel}
\be
  F=\frac{P_S}{P_S+P_T},
\ee
where $P_T=P_{T_{0}}+P_{T_{\up}}+P_{T_{\down}}$
and $P_S$ are occupation probabilities
for the triplet and singlet states, respectively.
The latter can be expressed by
$P_{\pm}=\bra{\pm} \hat{\rho}\ket{\pm}$ as
\be
   P_S=\left(\frac{\alpha}{\beta}+\frac{\beta}{\alpha}\right)^{-2}
   \left[\frac{P_{+}}{\beta^2}+\frac{P_{-}}{\alpha^2}\right],
\ee
with $\alpha=(1/\sqrt{2})\sqrt{1-\delta/(2\varepsilon_A)}$
and $\beta=(1/\sqrt{2})\sqrt{1+\delta/(2\varepsilon_A)}$.
Roughly speaking, when $P_S \gg P_T$,
fidelity reaches its maximal value $F\approx 1$,
while in the opposite situation it is suppressed.

Fidelity in the parallel magnetic configuration as a function of $\delta$ and $eV$
is shown in \fig{Fig:F} (a) in the absence of detuning, $\Delta\e = 0$, and (b) for $\Delta\e/U_{LR} = 0.4$.
One can clearly observe transport regimes where
$F$ is either equal to one or zero.
More specifically, for $eV<|(\delta+U_{LR})/2|$, one has $P_S \gg P_T$,
and fidelity reaches its maximal value with $F\approx 1$.
Thus, the transmitted pairs of electrons can be considered as entangled.
On the other hand, for bias voltages, $eV>|(\delta+U_{LR})/2|$,
the situation is just opposite and one obtains $F\approx 0$.
Consequently, the considered DQD setup
guarantees that, by properly tuning the device parameters,
fully entangled pairs of electrons can be extracted from
the superconductor and transmitted into normal leads.
This effect is insensitive to the value of the leads' spin polarization $p$
and the magnetic configuration of the system (results not shown).
It is also interesting to notice that fidelity provides
information about the flowing current.
The current does not flow due to the triplet blockade ($P_S=0$ and $P_T=1$),
i.e. when $F=0$, cf. Figs.~\ref{Fig:2} and \ref{Fig:F}.

The situation becomes more complex when finite detuning of DQD levels is present.
In this case the map of fidelity possesses richer structure, see \fig{Fig:F}(b).
The main difference is in the splitting of the line,
$eV\approx (\delta+U_{LR})/2$, along which the fidelity
varies between zero and one.
Now, one obtains $F=1$ for smaller bias voltages
for given detuning $\delta \gtrsim -U_{LR}/2$,
compared to the case of $\Delta\e = 0$.
In fact, the voltage is smaller by a factor of
level splitting $\Delta\e$.
Moreover, with increasing the bias voltage and for $\delta \gtrsim -U_{LR}/2$,
$F$ does not drop to zero immediately,
but becomes suppressed in a
nonmonotonic way in the transition region,
$eV\approx (\delta+U_{LR})/2 \pm \Delta\e$,
of width $2\Delta\e$, see \fig{Fig:F}(b).
Finally, we note that similar splitting
of the transition line separating the regions with $F=0$ and $F=1$
also occurs in the case of finite hopping between the dots
or finite magnetic field.

\section{Conclusions}

In this paper we have studied the spin-resolved Andreev transport
through double quantum dot-based Cooper pair splitters with
ferromagnetic leads. The considered device consisted of
two single-level quantum dots coupled to a common
$s$-wave superconductor and each dot coupled to
its own ferromagnetic lead. The calculations
were performed with the aid of the
real-time diagrammatic technique, assuming weak coupling
between DQD and ferromagnets and taking into account
the sequential tunneling processes.
We have analyzed the bias voltage and DQD level dependence of the
Andreev current and the differential conductance in the parallel
and antiparallel configurations, as well as the resulting
tunnel magnetoresistance.

In the case of infinite correlations in the dots, we have
discussed the behavior of spin-dependent characteristics in the
transport regime where only crossed Andreev
reflection processes are possible.
For certain DQD levels' configuration and applied bias voltage,
the current is then suppressed due to the triplet blockade.
We showed that even relatively large intradot correlations
can lead to finite leakage current in the triplet blockade.
We found an enhanced TMR in the triplet blockade, which
indicates the role of CAR processes in transport.
We have also analyzed the effect of splitting the
Andreev bound states by either finite DQD level detuning,
finite hopping between the dots or finite magnetic field.
While in the first two cases each Andreev bound state
becomes split into two, finite magnetic field further splits the ABS,
resulting in more complex transport characteristics,
with negative differential conductance and negative TMR occurring
in certain transport regimes.

Moreover, assuming finite correlations in the dots,
we have studied transport properties
in the full parameter space,
where both DAR and CAR processes are relevant.
We found transport regimes where
additional current suppression accompanied with
negative differential conductance occurs.
These suppression regimes are
due to enhanced occupation of certain many-body DQD states,
which diminishes the rate of either CAR or DAR processes,
depending on transport region.

Finally, in the CAR transport regime
we have also analyzed the entanglement fidelity
between electrons forming Cooper pairs.
We showed that the fidelity of split Cooper pair electrons can be tuned by
bias and gate voltages and for certain parameters $F$ can reach unity.
Consequently, DQD-based Cooper pair splitters, by properly tuning
the device parameters, can be sources of fully entangled pairs of electrons
that are extracted from superconductor and transmitted to normal leads.

\section*{Acknowledgments}

This work was supported by the National Science
Centre in Poland through the Project
No. DEC-2013/10/E/ST3/00213 and
Marie Curie FP7 Reintegration Grant No. CIG-303 689
within the 7th European Community Framework Programme.


\end{document}